\documentstyle[preprint,aps]{revtex}
\tighten
\begin{document}
\preprint{}
\draft
\title{Time resolved pattern evolution in a large aperture laser}
\author{
F. Encinas-Sanz, I. Leyva and J. M. Guerra\\ {\small\em Dept.
Optica, Facultad de CC. Fisicas\\ } {\small\em Universidad
Complutense de Madrid 28040 Madrid, Spain}}

\maketitle

\begin{abstract}

    We have measured quasi-instantaneous transverse patterns
 in a broad aperture laser. Non-ordered patterns yielding
 to boundary determined regular structures in progressive
 time-integrated recording are observed.
 The linear analysis and numerical integration of the full
 Maxwell-Bloch equations allow us to interpret the features of
 the experiment. We show that this system being far from threshold cannot be
 fully understood with a perturbative model.

\end{abstract}


%]%end of onecolumn

\footnotetext[1]{Research supported by the Spanish CICYT under
contract ..}

\newpage

Pattern formation in systems which exhibit spatio-temporal chaos
has been a field of intense research in last years. An averaging
process from chaotic to boundary-selected ordered patterns has
been observed in hydrodynamics \cite{NING93,GLUCKMAN93}. A similar
behaviour is foreseeable in other pattern forming systems, such as
large aperture lasers.

    In fact, this phenomena has been predicted for lasers from the
basis of the Maxwell-Bloch equations
\cite{HUYET1,HUYET3,STALIUNAS}. However, due to its extremely fast
evolution, the time resolved spatio-temporal dynamics of a broad
area laser has never been observed. So far the experimental work
only analyze averaged patterns, mainly in cw CO$_2$
\cite{HUYET1,HUYET2,LABATE,DANGOISSE} and semiconductor lasers
\cite{HEGARTY99}, since the minimum reachable exposure time was
about 1$\mu s$, too long to obtain information about the pattern
history.

    In this letter we study, both experimentally and theoretically, the time
  resolved dynamics of a large Fresnel number pulsed laser. The measurements
  were carried out with a system that we have recently developed to obtain
  infrared snapshots with a short exposure time (minimum $\simeq$ 1 ns).
  This setup has been described in detail in Ref. \cite{OSCAR},
  and allows us to record virtually instantaneous transverse intensity laser patterns.

   The source was a transversely excited atmospheric (TEA) CO$_2$ laser,
  with Fresnel number ${\cal F} =\frac{b^2}{\lambda L}\simeq 10$, where
  $2b=20$ mm is the laser aperture, $L=1$ m the resonator length and $\lambda
  =10.6\times10^{-6}$m the lasing wavelength. This laser emits about 15$\%$
  of the output energy in a gain switch pulse ($\sim$
70 ns), followed by a long collisional transfer tail ($\sim$ 2-3
$\mu$s) \cite{FER3}. Being a pulsed laser, it offers interesting
characteristics such as high pumping far from threshold and wider
detuning, which allow the exploration of additional features of
the class B laser behaviour.

In order to study the dynamics of the system, we take snapshots
all along the duration of the laser pulse. A sample of recorded
patterns taken with a 6 ns temporal resolution at different times
along the pulse is shown in Fig. \ref{f:PATTERN}. Note that all
the instantaneous patterns are disordered and are
 non-reproducible from shot to shot (Fig. \ref{f:PATTERN}a, b, c, (color)).

 However, such irregular appearance masks some kind of regularity, since the
 averaged patterns integrated all along the pulse length are ordered and reproducible
 (Fig. \ref{f:PATTERN}d), having eight or nine
 rolls parallel to the almost flat laser electrodes. The transverse spatial
 period of those bands is $\simeq$ 1.8 mm, similar to the size of the intensity
 maxima appearing in the disordered patterns. No sign of two spatial structure
 scales can be observed, contrary to the predictions of Ref. \cite{HUYET1}. It is remarkable
 that the same regular structure is recovered by averaging over many equivalent
 instantaneous patterns, observation that was suggested by the results in
 Ref. \cite{FER1}. In this sense, a sort of ergodicity is observed.

We have also studied the temporal evolution of the intensity in
one small area ($\simeq$ 1 mm$^2$), for which a photon-drag
detector was used (rise time $\simeq$ 1 ns). We find that the
local intensity oscillates in a completely irregular form (Fig.
 \ref{f:TIME}a), with a period of 10 ns approximately. Furthermore,
  the cross correlation between the local oscillation measured at
  two different points of the patterns is
very low, even if they are as close as 6 mm\cite{PASTOR}.

 As the characteristic period of the intensity local fluctuations is about 10
 ns, in order to record true instantaneous snapshots
it would be convenient to reduce the width of the temporal
 window as much as possible below this value. But, if the window width
 is too small we do not get enough photons to be recorded by our system.
 On the other hand, a snapshot
 recorded with a much longer exposure integrate the pattern over several periods and
it cannot be considered instantaneous any longer. That justify the
6 ns choice that we have made for the temporal window width.

Nevertheless, it would be interesting to know how much recording
integration time is necessary to obtain a sufficiently ordered
pattern. Therefore, in order to follow the averaging process, we
have also measured sequences of progressively integrated patterns
by varying the temporal window (Fig. \ref{f:GS}). These
measurements were made in the first 150 ns of the laser pulse, in
which the intensity is several times larger than in the pulse
tail, and therefore the limitation due to the small intensity
disappears. That allows us to reduce the time window down to 2 ns
(Fig. \ref{f:GS}a), showing a small number of intensity maxima
which contrast deeply with the background. We see that the longer
exposure, the bigger number of maxima appear in the (Fig.
\ref{f:GS}b, c), approaching progressively the ordered pattern,
whose rolls are already clearly recognizable (Fig. \ref{f:GS}d)
when the time integration is about 100 ns, yet much shorter than
the total pulse duration.

 Our theoretical approach to the problem includes both numerical
 simulations and a linear stability analysis of the full
 Maxwell-Bloch laser equations.

Hence, in order to reproduce the spatio-temporal dynamics, we
directly integrate the two-level Maxwell-Bloch equations
\cite{HUYET3}:

\begin{eqnarray}
 \frac{\partial E}{\partial t}
&=&-\kappa[(1-i\delta)-i\frac{a}{2}\triangle_t]E-\kappa QP\; ,\\
 \frac{\partial P}{\partial t}&=&-\gamma_\perp[DE+(1+i\delta)P]\; ,\\
 \frac{\partial D}{\partial
 t}&=&-\gamma_\parallel[D-1-\frac{1}{2}(EP^*+E^*P)]\; ,
\label{MB}
 \end{eqnarray}
where E=E({\bf x},t) is the slowly varying electric field,
P=P({\bf x},t) the polarization, D=D({\bf x},t) the population
inversion, Q=Q({\bf x},t) the rescaled pump,
 $\kappa=\frac{-c}{2L}{\rm ln}(R)$ representing the cavity
 losses with $R=\sqrt{R_1R_2}=0.78$ the resonator reflectivity
 , $a=\frac{c}{2 \kappa L{\cal F}}$ a diffraction coefficient,
$\delta=\frac{\omega_{21}-\omega}{\gamma_\perp}$ the rescaled
detuning and $\triangle_t$ the Laplacian in the adimensional
 transverse coordinates of the system {\bf x}=(x,y).
 In an atmospheric laser, the decay rates can be chosen as
  $\gamma_\perp = 3 \times 10^9$ s$^{-1}$, $\gamma_\parallel = 10^7$s$^{-1}$
  and $\kappa = 3.9 \times 10^7$ s$^{-1}$.

 The transversal pumping profile is taken to be homogeneously distributed along
 one transverse  axis and Gaussian in the other, in order to
reproduce the experimental current spatial distribution of the
discharge. Likewise, the temporal form of the pumping was
simulated by a function approximating the pulsed excitation,
 with typically large pumping parameters in the maximum (Q$_{max}\simeq$15-25).
 Null boundary conditions, equivalent to the experimental ones, were used.

The patterns and local intensity temporal evolution obtained by a
standard numerical method, look very similar to those recorded
experimentally. In the spatial domain, the instantaneous patterns
(Fig. \ref{f:PATTERN}e) are well reproduced, as well as the time
integrated (Fig. \ref{f:PATTERN}f). In both them, only one spatial
scale of structures is found, in accordance with the experimental
records. The numerical temporal evolution (Fig. \ref{f:TIME}b),
although completely irregular, presents a characteristic time of
the same order than its experimental counterpart (Fig.
\ref{f:TIME}a).

As it has been mentioned, this time scale of the irregular
oscillations measured in the local dynamics is around a few
nanoseconds (Fig. \ref{f:TIME}), whereas the typical evolution
time of the pulsed pumping is a few microseconds (2-3 $\mu$s). A
good approximation is to consider the dynamics measured in the
slow varying pulse tail to be quasi-stationary  \cite{HARRI}.
Thus, we can still use the properties of a linear stability
analysis to gain some deeper insight into the observed irregular
dynamics. Taking this consideration into account, we undertake the
stability study linearizing Eqs.(1-3). As it is known, the
Liapunov exponents of the equations are the real parts of the five
roots of the corresponding secular quintic equation, which cannot
be solved analitically, and therefore it is not possible to obtain
algebraic expressions for the eigenvalues. However, it is known
that, for class B lasers, out of the five roots, one is real and
the rest come in complex conjugated pairs \cite{HUYET3,HUYET2}.

 It can be shown that the resonator can develop the instability
 associated to the real root only when the condition

\begin{equation}
\delta \frac{Q-(1+\delta^2)}{Q}\gg \frac{-\pi}{4{\cal F} {\rm
ln}(R)}\; , \label{deltacond}
\end{equation}
is satisfied \cite{HUYET3}.  In a TEA-CO$_2$ laser, only one
molecular transition oscillates (P20 line,
 Ref.\cite{VOROBEVA}). In the present case, it includes a large number of axial modes
 simultaneously (15-20), with a free spectral range of $\frac{c}{2L}\simeq $150 MHz
 \cite{PASTOR}. Hence, most of the values of $\delta$ lie in the
 interval $ -1\leq \delta\leq 1$. Since $ Q \gg 1$, the condition
 (\ref{deltacond}) is not satisfied for the most
of the oscillating axial modes. Thus, we conclude that this
instability does not actually affect the laser, or does so very
weakly.

 In addition, two of the complex conjugate roots can be approximated as
$-\gamma_\perp(1\pm i\delta)$, being their real part always
negative and therefore not associated to any instability.

 The detuning value determines whether the real parts of the two remaining
 conjugate roots are negative for every transverse wave-vector $k$, or
 positive for an interval around a value $k_o(\delta)$, the wavelength vector at
 which the Liapunov exponent is maximum and positive (Fig. \ref{f:STAB}).
 Each set of system parameters has a certain critical detuning
 $\delta_c$, such as for $\delta>\delta_c$, there is an interval
  of wave numbers $k$ for which the real part of the root is positive.
 In the present case this critical value is rather low ($ \delta_c \sim 0.06$).
 Then, for the most the axial modes with $\delta > 0 $, the instability
 associated to those complex conjugate roots shows up.

 Summarizing, in this kind of laser most of the axial
 modes with positive detuning bear the short-wavelength instability due to the
 pair of complex conjugate roots, whereas that associated to the real root is
 not supported because of the diffraction (Fig.\ref{f:STAB}). Therefore, the observed
 irregular spatio-temporal behaviour can only be justified by the action of
 the remaining instability, in contrast with \cite{HUYET2}.

 On this basis, it is possible to estimate the expected spatial
 and temporal scales. The spatial scale of the instabilities
 associated to the complex root will be around $k_{o}^{-1}$.
 By solving numerically the secular equation with $\delta$ = 0.6 and
 $Q = 6.0 $ (a reasonable mean
value in the pulse tail), we obtain $k_{o}= 3225$ m$^{-1}$. Then
the size of the generated structures should be¡

\begin{equation}
S_0= 2\pi k_0^{-1} \simeq 1.94 \times 10^{-3} \;\;{\rm m}
\label{size0}
\end{equation}

The average size of the experimentally measured intensity maxima
in the instantaneous
 patterns, and consequently the spatial period of the bands appearing in the time
 integrated patterns is

\begin{equation}
 S_{exp}=1.8\times 10^{-3} \;\; {\rm m}
\label{sizexp}
\end{equation}

The agreement between (\ref{size0}) and  (\ref{sizexp}) relates the observed
 dynamics with the remaining instability.

Furthermore, the imaginary part of these eigenvalues gives the
oscillation frequency of the solutions. Hence, for the same $k_o$,
we obtain $\omega\simeq 360 \times 10^6 s^{-1}$, corresponding to
a period of 16 ns, very close to both the experimental and the
numerically obtained fluctuation period of $\sim$ 10 ns (Fig.
\ref{f:TIME}a, b). Thus, the time scale is also well predicted by
the stability analysis.

 Concerning the physical interpretation of the irregular observed dynamics,
  in the present case the suppression of the instability coming from
the real root invalidates its being considered as the origin of
the disordered instantaneous pattern. Thus we found that even
though the distance to the threshold is moderate here ($Q\sim 6$),
an analysis based on the order parameter equations seem to be
already insufficient. To test this, the same numerical integration
has been carried out for an hypothetical case nearer threshold
($Q_{max}=2$), where the order parameter equations must dominate
the amplitude behaviour \cite{HUYET3}. In agreement with this
perturvative approach, since the phase instability is inhibited
here, the instantaneous patterns show a large degree of order
(regular rolls which oscillate periodically, Fig.
\ref{f:THRESHOLD}). However, in the far threshold pumping case
they are disordered, and therefore, the present problem does not
seem suitable for reduction to a perturbative one, as is usual in
the theoretical approaches. In other words, in the observed
dynamic it is not possible to distinguish between phase and
amplitude fluctuations. This result is a test of the validity
range of the order parameter equations, which was not easily
verified experimentally. As a more probable origin of the
phenomenon, a secondary instability of the traveling wave
solutions can be suggested \cite{CROSS}.

In conclusion, in this work we report the measurement of time
resolved intensity patterns in a large aperture laser, by means of
an experimental system developed in our laboratory. A rich
irregular intensity spatio-temporal dynamics, usually masked under
time integrated measurements, has been uncovered. We show
experimental evidence of how this local irregular dynamics
averages to boundary determined order, as had been observed in
other pattern forming systems but so far only predicted
theoretically for lasers. Besides, a numerical integration of the
two-level full Maxwell-Bloch equations and its corresponding
stability analysis reproduce the experimental observations with
outstanding agreement.

\begin{figure}
  \caption{ Instantaneous patterns at different times of the
laser pulse.
 Recordered: a)150 ns b)300 ns c)500 ns delay from the gain-switch pulse.
  Numerically generated: e)500 ns delay with $Q_{max}$=18.0 and $\delta=0.6$].
  Time-integrated patterns [experimental, d), and numerical f)].
  Experimental pattern dimension 20$\times$20 mm.}
  \label{f:PATTERN}
\end{figure}

\begin{figure}
  \caption{Time evolution of the local intensity, a) measured
b) numerically generated for $Q_{max}$=18.0,$\delta=0.6$. }
 \label{f:TIME}
\end{figure}

\begin{figure}
  \caption{Patterns recordered with different exposure time: a)2 ns ,
  b) 6 ns c)30 ns d) 100 ns}
  \label{f:GS}
\end{figure}

\begin{figure}
  \caption{ Real($\lambda$) for several $\delta$ values,
$K_{lim}$  being the diffraction limit for the instabilities, and
$K_{exp}$ the experimentally found mean wave number.}
  \label{f:STAB}
\end{figure}

\begin{figure}
  \caption{ Numerically generated instantaneous pattern for $Q_{max}$=2.0, $\delta$=0.6.}
  \label{f:THRESHOLD}
\end{figure}

%\pagebreak

\end{document}